# Scientific Understanding and the Risk from Extreme Space Weather

M.A. Hapgood, Science & Technology Facilities Council, Rutherford Appleton Laboratory, Harwell Science and Innovation Campus, Didcot, Oxfordshire, OX11 0QX, United Kingdom

## Abstract

Like all natural hazards, space weather exhibits occasional extreme events over timescales of decades to centuries. Historical events provoked much interest, and sometimes alarm, because bright aurora becomes visible at mid-latitudes. However, they had little economic impact because the major technologies of those eras were not sensitive to space weather. This is no longer true. The widespread adoption of advanced technological infrastructures over the past fifty years has created significant sensitivity. So these events now have the potential to disrupt those infrastructures – and thus have profound economic and societal impact. However, like all extreme hazards, such events are rare, so we have limited data on which to build our understanding of the events. This limitation is uniquely serious for space weather since it is a global phenomenon. Many other natural hazards (e.g. flash floods) are highly localised, so statistically significant datasets can be assembled by combining data from independent instances of the hazard recorded over a few decades. Such datasets are the foundation on which reliable risk assessment methodologies are built. But we have a single instance of space weather so we would have to make observations for many centuries in order to build a statistically significant dataset. We show that it is not practicable to assess the risk from extreme events using simple statistical methods. Instead we must exploit our knowledge of solar-terrestrial physics to find other ways to assess these risks. We discuss three alternative approaches: (a) use of proxy data, (b) studies of other solar systems, and (c) use of physics-based modelling. We note that the proxy data approach is already well-established as a technique for assessing the long-term risk from radiation storms, but does not yet provide any means to assess the risk from severe geomagnetic storms.  This latter risk is more suited to the other approaches, but significant research is needed to make progress. We need to develop and expand techniques to monitoring key space weather features in other solar systems (stellar flares, radio emissions from planetary aurorae). And to make progress in modelling severe space weather, we need to focus on the physics that controls severe geomagnetic storms, e.g. how can dayside and tail reconnection be modulated to expand the region of open flux to envelop mid-latitudes?

# 1 Background

Over the past decade space weather has developed strongly as a scientific discipline that can help to address the challenges that the space environment poses to our technological civilization. The impact of space weather on technology has been known since the mid-nineteenth century, e.g. through the impact of geomagnetically-induced currents on telegraph systems. However, our scientific understanding of the processes that cause these impacts has emerged only in the past few decades, mainly through our growing appreciation of the plasma processes that dominate the space environment (e.g. the nature of the solar wind and the role of magnetic reconnection).

In parallel with these scientific developments, our civilization has greatly increased its sensitivity to space weather over the past forty years as technological systems have become more critical to the functioning of economies and societies around the world. In particular two particular sensitivities stand out as critical at this time.

1. Our dependence on space-based systems, both public and commercial, for a range of critical services including communications, navigation, meteorology, environment and security monitoring. It is now widely recognized that these systems are at risk from the natural space environment as well as human factors – and that the assessment and mitigation of that risk should be a matter of public policy because of the societal importance of these systems. The policy response in the US, and now in Europe, has been to establish space situational awareness (SSA) programmes. These programmes seek to develop services that provide public and private sector users with awareness of the space environment and its impact on their activities, e.g. operation and exploitation of space-based systems. Space weather is a key element in the SSA programmes through its effects on spacecraft (e.g. radiation, charging, drag) and on their radio links (e.g. phase shifts and scintillation arising from transmission through the ionosphere and plasmasphere).
2. Our deep dependence on electrical power. This has gathered pace over the last forty years, e.g. with the greatly reduced use of coal as a fuel for railways, factories and domestic heating and the use of large-scale power grids to deliver power from low-cost sources distant from consumers. Space weather can interfere with this through the production of geomagnetically induced currents in the solid body of the Earth. These can enter power grids through earth connections and reduce delivered power by unbalancing grid operation. In the worst cases, they can degrade and destroy devices such transformers through vibration and heating. The prime example of this hazard is the power grid failure in Quebec in 1989. This acted as a wake-up call and the power industry in regions of high risk (e.g. North America) has worked to reduce the hazard by maintaining awareness of the threat and taking procedural action, e.g. reducing long-distance transmission of power during periods of risk. Nonetheless the space weather threat is still significant, e.g. through extreme events beyond the capability of existing mitigation procedures and through the lack of awareness when new power grids are developed.

Other critical sensitivities are likely to appear in the future as new technologies are exposed to space weather. One emerging candidate is the potential of solar energetic particle events to disrupt digital systems on the ground. Chip vendors now advise that cosmic radiation is the main source of error in digital systems and that design of critical systems must allow for this. Major SEP events are known to increase ground-level radiation many-fold – the largest known case being factor 50 (Marsden et al., 1956; Gold and Palmer, 1956). Another candidate is the possibility that bursts of intense radio noise from the Sun could interfere with many modern wireless technologies (e.g. mobile/cell phones, wireless control systems, wireless internet). The latter is very similar to another early impact of space weather: UK scientists working on radar systems during the Second World War discovered that mysterious interference in those systems originated from solar radio emissions and not, as they had first feared, from hostile action (Hey, 1946, Lanzerotti et al, 2005).

Space weather, like the normal weather in the Earth's troposphere, is episodic. It experiences long periods of quiet conditions when impacts on technological systems are small interspersed with periods of disturbed conditions which have modest impacts. Extreme events are rare but do happen and it is during these events that the sensitivity of technological systems will be most starkly revealed. A key issue therefore is to estimate what is the extreme risk that society should plan for: what is the frequency occurrence of extreme space weather events? - and what are the potential consequences of such events? The paper outlines the current state of knowledge on such events and argues that further scientific research is critical to establishing an adequate knowledge base on extreme space weather. It also proposes some possible approaches to that future research and explores their strengths and weaknesses.

## 2 Space weather as a natural hazard

This paper treats space weather as an emerging natural hazard. This is an important step in distinguishing the study of space weather from the more general study of the space environment (e.g. as solar and solar-terrestrial physics). A key issue in defining natural hazards is the interaction of the natural environment with human beings and the infrastructures that they have built. To qualify as a natural hazard an environmental event must have the potential to affect people or their activities; an event that does not affect people is purely of academic interest and is not a hazard.

The risk from space weather is largely focused on the reliable and safe operation of technologies that enable human activities. There are only limited direct threats to human health – mainly through radiation effects and even then only because technology (aviation and spaceflight) allows humans to travel in regions where there is an enhanced risk from cosmic radiation. There is some scientific speculation that space weather can have other effects on human health (Palmer et al, 2006) but this is remains an intriguing idea for which this is no scientific consensus. The major impacts of space weather are those on human technologies. These have gradually emerged over the past 150 years starting with the development of telegraph and its vulnerability to geomagnetically-induced currents. Space weather impacts have gathered pace over the past fifty years, e.g. through the deployment of critical infrastructures in space, through the ubiquitous use of electrical power and radio-based systems and, most recently, the embedding of software controls in a vast range of technologies. It is this growth in the use of vulnerable technologies that marks space weather as an emerging natural hazard that should be considered alongside more traditional examples.

Those traditional hazards illustrate the concept of natural hazards. They typically arise when adverse conditions occur as occasional events in environments that are generally benign. For examples there is often a physical correlation between the general fertility of a region and the risk of occasional dangerous events. Examples include (a) river valleys where the availability of water for agriculture is correlated with the risk of flooding, (b) sea coasts where the availability of seafood should be balanced against the risk of flooding by storm surges and by tsunamis, and (c) the margins of volcanoes, where the high soil fertility is correlated with the risk from lava and ash. Natural hazards occur in places in that are good to live, except on those few occasions when conditions deviate far from the norm.

Space weather is similar in that humankind has necessarily evolved on a planet that is favourable to the development of life. The Earth sits at a comfortable distance from the Sun in what is now termed the "Habitable zone" by astronomers studying the potential for life on exo-planets. A planet orbiting in this zone is likely to have surface temperatures that allow the presence of liquid water, an important factor for life. However, the potential for life also depends on several factors increasingly familiar to the solar-planetary physics community, and now gaining visibility in exo-planet studies (Khodachenko, et al, 2009).

- The need for a benign radiation environment such that living organisms can reproduce with only minor chance of mutation from generation to generation. Life on Earth is protected from the severity of cosmic radiation by a number of factors: (a) the scattering of galactic cosmic rays by magnetic structures in the solar wind, (b) the deflection of solar energetic particles and low energy (<10 GeV) cosmic rays by the geomagnetic field and (c) the absorption of the latter particles by the relatively thick so atmosphere. These factors all work together such that cosmic radiation provides only 10% of the natural radiation exposure at sea level on Earth and becomes the dominant source only at high altitudes (>3000 m above sea level).
- The need to protect planetary atmosphere from erosion by the stellar wind. Over billions of years this can provide sufficient momentum to remove the atmosphere – but only if that momentum can be coupled into the planetary atmosphere. On Earth the geomagnetic field creates the magnetosphere, a diamagnetic cavity that holds the solar wind away from the bulk of the atmosphere. Some momentum coupling occurs as a result of magnetopause reconnection and ion-neutral coupling in the high latitude ionosphere – and significant outflows have been reported from this region (Lockwood et al, 1985, Ogawa et al, 2003, Engwall et al., 2009).

But these factors that give Earth its potential for life are also ones that expose it to space weather hazards. Its relative closeness to the Sun is essential for life, but also means that it is exposed to the environmental effects that constitute space weather. For example, the solar X-rays and UV fluxes that drive many ionospheric space weather effects follow the inverse square law, so their variations at Earth will be tens or hundreds of times greater than amongst the outer planets such as Jupiter and Neptune. Similar considerations apply to other space weather drivers such as the energy and momentum fluxes in the solar wind and also to the fluxes of solar energetic particles (though the latter's radial variation will modified by particle scattering and shock acceleration at structures away from the Sun). Nonetheless it is clear that Earth's closeness to the Sun is one factor in creating the risk from space weather.

Another factor that raises the space weather hazard at Earth is the magnetosphere. As discussed above, this can act as a shield protecting much of the Earth from the solar wind and solar energetic particles. But it can also act as a solar wind energy collector - when magnetopause reconnection occurs and thus allows solar wind energy to cross the magnetopause. Since the cross-section of the magnetosphere is hundreds of times greater than that of the Earth itself, so this process can collect much more solar wind energy (compared to an unmagnetised planet) and focus it in the polar regions (whereas the energy input would spread over the whole dayside hemisphere in the case of an unmagnetised planet). A key consequence of this focused energy input is heating the polar upper atmosphere, which has profound impacts on atmospheric behaviour at all latitudes: (a) it can reverse the pattern of upper atmosphere winds so that they flow away from the heated polar regions rather than the sub-solar region of the dayside, (b) it can change the composition of the upper atmosphere through enhanced vertical transport, and (c) variations in the polar energy input can generate large-scale acoustic gravity waves that propagate to lower latitudes. These changes in atmospheric behaviour give rise to well-known space weather problems such as changes in atmospheric drag on spacecraft and a wide range of changes in ionospheric behaviour. The solar wind energy input also generates the aurora and associated space weather effects such as spacecraft charging in low polar orbits, enhanced ionospheric scintillation at high latitudes, high variability in HF radio propagation conditions on trans-polar routes and geomagnetically induced currents (GIC) in power grids, pipelines and signalling systems.

## 3  Some examples of severe space weather events

In this section we present some examples of severe space weather events to illustrate what we know of their impacts and what we know of the physics of such events. These examples are not intended to be a complete catalogue but rather a subset that illustrates the extent of our knowledge. To rank the severity of geomagnetic activity in these events we use the aa*MAX index developed at the US National Geophysical Data Center. This index is based on the well-known aa geomagnetic index, but uses a running mean technique to provide a robust estimate of the overall strength of any storm. For more details of this index see Appendix A.

We first consider a number of major events recorded over the past forty years a period for which there is reasonable data coverage by both ground-based and space-based sensors:

*13 March 1989.* This is the most notable event of the past forty years. It has, by far, the largest value of aa*MAX (441) in that period. Indeed it ranks highest in the whole series of aa*MAX values. The next strongest events are those of 18 September 1941 (aa*MAX

= 429) and 23 March 1940 (aa*MAX = 377). However, we have very limited information on the space weather impacts arising during those earlier events. Much of the present technological sensitivity to space weather did not then exist – and what interest did exist was often focused on specialist niches such as radio communications. In contrast, the 1989 event attracted much interest because of its space weather impacts, so there is substantial information on these. Most notable was the power failure in Quebec, which acted as a major wake-up call for those concerned with space weather effects on power grids. Since that time, much has been done to mitigate the risk, e.g. through improved operational procedures in areas known to be at risk such New England. The 1989 event was also notable for the changes in atmospheric drag, which caused the US space surveillance activities (then operated as NORAD) to loose track of over 1600 space objects. The March 1989 was less notable for its enhanced radiation fluxes and were over-shadowed by major solar radiation storms in October 1989, which did not generate a large magnetic storm.

*29 October 2003*. This event marked the peak of the so-called Halloween storms of 2003. It has aa*MAX = 332, the second highest value of the past 40 years and 9$^{th}$ highest since records started in 1868. The Halloween storms have attracted much attention because they are the strongest storms observed with the comprehensive space-based measurements that emerged in the 1990s under the auspices of the International Solar-Terrestrial Physics programme (Whipple and Lancaster, 1995). A wide range of space weather impacts were reported. There were many problems with spacecraft operations, e.g. control anomalies due to single event effects, loss of data due to space weather interference with sensors. An interesting example was the loss of stable magnetic orientation in geosynchronous orbit and thus the need to disable spacecraft sub-systems that rely on that stable field. There was also considerable space weather interference with high-frequency radio communications, especially in polar regions. This required the re-routing of trans-polar flights in the Arctic and use of backup communications for groups working in the Antarctic. Interference with radio communications was also reported in other frequency ranges – typically when activity occurred on the Sun while it was aligned with the main beam or major side lobes of the receiving antenna. The impact on power grids was noticeably less than in 1989, especially in North America where operators applied the lessons learned in 1989. Some problems were reported in Northern Europe and South Africa, in particular overheating of transformers. One special feature of this event was the occurrence of significant space weather effects on space missions well away from Earth, e.g. the failure of the radiation monitor on NASA's Mars Odyssey mission and severe interference in the star trackers on ESA's Mars Express mission.

*8 February 1986*. This event ranks 20$^{th}$ highest in the record of aa*MAX values from 1868, and fourth highest of the past forty years, with aa*MAX = 287. The key lesson learned about this event was its occurrence close to the solar minimum (which occurred in March 1986), providing a clear demonstration that the solar cycle has only a statistical influence on the frequency of severe events; they are rare during solar minimum, but not impossible.

Unfortunately these three recent events provide only limited scientific insights into what makes a severe space weather event. All three events were reasonably well-observed by ground-based sensors such magnetometer networks. To give one example, Figure 1 shows the variation of the four auroral electrojet indices during the 1989 event. The shape of the variations is not unusual and a series of perhaps ten sub-storms can be seen following the sudden commencement at 01:28 on 13 March and ending as activity died down on late on 14 March. What is unusual in this plot is the scale – an order of magnitude larger than usually used to plot these indices. This demonstrates the intensity of the event. The auroral electrojet indices for the 1986 and 2003 events exhibit similar behaviour.

Unfortunately none of these events were well-observed by space-borne sensors. In particular, there is limited information on the variations of the solar wind and the heliospheric magnetic field during the event. As discussed above the 1986 and 1989 events pre-date the current era of comprehensive space measurements. At that the only spacecraft that made routine measurements of solar wind and heliospheric magnetic field was NASA IMP-J spacecraft – and these measurements were constrained by the spacecraft orbit and limited downlink capacity. As a result there are no useful data from the 1989 event. There are some data for the 1986 event but these are sparse and disjointed so it is very difficult to deduce what happened. The 2003 measurements were limited by a different factor. In this case, there was good availability of suitable instruments and downlink, but the solar wind sensors were saturated by the high fluxes of solar energetic particles. Thus the state of the solar wind and its coupling into the magnetosphere is not well known in any of these events. This is a major obstacle to exploring the physics of severe space weather events.

We now consider two older events, which provide evidence of stronger space weather effects:

*23 February 1956*. This event exhibited the one of the most intense radiation fluxes that has been observed instrumentally. As it precedes the advent of spaceflight, the effects were only observed using ground-based instruments. Fortunately, the importance of those observations was recognised at the time. They were collated and published as a set of short papers by Gold and his colleagues (e.g. Gold and Palmer, 1956; Marsden et al, 1956). Those observations show an enormous ground-level cosmic ray event for a short period, the ground level radiation fluxes were enhanced fifty-fold, which remains the largest enhancement ever observed instrumentally. Subsequent analysis of these observations suggests that this event was distinguished from other recent radiation storms by the hardness of the particle spectrum. The event included significant particles fluxes at energies above 1 GeV and thus was capable to delivering significant neutron fluxes to ground-level. The event has been widely used as an example of the radiation threat to avionics, e.g. Dyer has shown that a repeat could generate single event upset rates greater than 1 per minute at 10 km altitude and 1 GeV rigidity cut-off (Dyer, 2002). This would put most transatlantic air routes between Europe and North America at severe risk.

This event is an example of an intense radiation storm that was not associated with strong geomagnetic activity. There was no elevation of geomagnetic activity during the storm, but it was followed by modest storm that started during the following day. This had aa*MAX=131, which ranks 165$^{th}$ in the list of storms.

It is also an example of a high flux radiation event with very short duration. As a result the fluence (i.e. time-integrated flux) was well below that recorded in major radiation storms that have been observed over the past seventy years. It is an event that emphasises space weather risks that arise from high fluxes, e.g. single event effects in critical electronic systems. These must be distinguished from effects (e.g. cancer risk for astronauts) that are linked to total radiation dose and thus are sensitive to fluence rather than flux.

*1 and 2 September 1859.* This is the outstanding space weather event on which we have any significant data and was the climax of an extended period of space weather activity from 28 August to 9 September 1859. It is often termed the Carrington event in honour of Richard Carrington, one of two astronomers who observed the solar flare that initiated the event (Carrington, 1859; Hodgson, 1859). These observations marked the discovery of solar flares. A huge magnetic storm started some 17 hours after the flare and continued for some 36 to 48 hours. This suggests that the solar ejecta (coronal mass ejection) that powered the storm travelled at a very high speed (~2400 km s$^{-1}$). During the storm, bright aurorae were observed around the world and reaching down to geomagnetic latitudes as low as 20° (Green and Boardsen, 2006). The storm induced strong currents in telegraph systems, which were advanced communication system of the time. By 1859 land-based telegraph systems had been extensively deployed in Europe and North America and these experienced many problems during the storm – see Boteler (2006) for an extensive discussion of this issue. We also have a useful set of magnetometer data from 1859. Observations made at Kew in London showed that there was sharp 100 nT magnetic pulse at the time of the flare (e.g. see Boteler (2006) for a reproduction of the Kew magnetometer recordings for the whole storm period). We now recognize this pulse as a "magnetic crochet" arising when the flare X-rays enhance the plasma density at altitudes around 100 km. At this altitude the ion dynamics are dominated by collisions with neutrals but the electron are collisionless. As a result the plasma generates a significant cross-field current – and it is this current, enhanced by the flare-induced ionization, that produces the magnetic crochet. Another valuable magnetometer observation from the Carrington event is the large depression (~1600 nT) in the field observed at Mumbai in India (Tsurutani et al, 2003). This is a huge perturbation for station at such low geomagnetic latitudes – almost three times larger than seen at any other low latitude station and at any other time. Another piece of evidence has emerged from proxy studies, specifically studies of gases trapped in ice-cores. These show that a large excess of nitrates was trapped in the ice layer deposited in 1859, strongly suggesting that there was a massive production of nitrates in Earth's atmosphere around the time of the Carrington space weather event (Shea et al, 2006). The ice core proxy data suggests that the Carrington event had the largest fluence of any solar radiation storm of the past 450 years.

The Carrington event is now widely used as the canonical example of extreme space weather. It is a clear example of a very extreme event - in terms of the energetic particle radiation reaching the Earth, in terms of the short travel time of the solar ejecta that caused the magnetic storm and in the scale of that storm as evidenced by auroral observations to very low geomagnetic latitudes. The space weather impact of the event was modest – mainly appearing through disruption to telegraph services, which were then becoming an important element in medium-distance communications, e.g. between cities and across railway networks. The use of electric telegraph for global communications was still a decade or two in the future, so the storm had little impact on international activities. The other major technologies of that era were not sensitive to space weather. These include:
- Increasing use of steam power for major transport systems with wind power still a major factor for maritime transport
- Maritime navigation based on optical and mechanical devices such as sextants and chronometers
- Universal use of draft animals for short range transport
- Communication by written documents carried by the above transport systems
- Well-established use of steam power for industrial machinery but with water and wind power still used in some applications such as milling of grain and cloth
- Intensive use of human physical effort across industry
- Use of coal for heating of buildings
- Use of gas derived from coal for lighting
- Optical telegraph systems for high-speed communications on critical routes– though these were gradually being superseded by the electric telegraph

Thus the world of 1859 was dominated by a range of technologies that, with the exception of the electric telegraph, were insensitive to space weather. As a result the Carrington event was no more than a fascinating nuisance to that world.

The situation today is very different. The technologies of 1859 have gradually been retired and replaced by technologies that have improved the standard of living around the world – with much of the change taking place over the past forty to fifty years, e.g. the move away from steam power in transport systems. Unfortunately, many of these technologies have brought a greater sensitivity to space weather especially through use of electrical power and space-based infrastructure. There is significant evidence that a repeat of the Carrington event would challenge operation of both power grids and spacecraft.

The challenge to power grids would come if the event were to generate geomagnetically induced currents over a wide range of latitudes, i.e. beyond the sub-auroral latitudes where the worst problems have previously been recorded. These effects could extend to lower latitudes where there are large populations and thus widespread societal dependence on the reliable operation of electrical power grids. For example, it has been estimated that a repeat of the Carrington event could severely damage the US power grid with adverse economic impact in the range of one to two trillion dollars (Space Studies Board, 2008).

The sensitivity of power grids to severe space weather may further increase as we seek to exploit renewable sources of electrical power (solar, wind, tides and hydro) which may be located far from population centres. For example, it has been proposed that the future architecture of power generation in Europe should exploit solar power systems based in Southern Europe and North Africa, plus wind and tidal power systems based on the Atlantic margin. This architecture is attractive as it would provide large scale power generation while addressing major concerns such climate change and security of supply. But it would require long distance transmission of electrical power and thus could be vulnerable if a severe space weather storm were to induce currents across Europe and down in North Africa. It is therefore important that new power architectures are designed to mitigate severe space weather events.

The challenge to spacecraft (and thus to space-based infrastructure for communications, navigation and earth surveillance) would come from the severe radiation environment. This would have the potential to disrupt and even destroy a large part of the operational spacecraft fleet, especially through single events effects such as bit-flipping and latch-up and through degradation of solar power arrays. The impact on the global spacecraft fleet has been modelled by Odenwald et al (2006). They estimated the global economic impact at 44 billion dollars in terms of loss of income from space-based services and 24 billion dollars in terms of spacecraft losses.

The Carrington event is a credible example of a severe space weather event. We know in outline what happened on 1 and 2 September 1859 and our knowledge of the underlying physics tells us that a repeat will happen one day. Indeed, there is no scientific reason to exclude the possibility that a bigger event will occur one day. What we don't know with any certainty is the likelihood of a repeat within, say, the next century, let alone the next thousand years. This is an open issue that needs further research.

In the mean time we must use the Carrington event as our canonical example of a severe space weather event. It provides concrete evidence that the Sun can occasionally produce events that will challenge key technologies that have emerged over the past fifty years and that are now critical underpinning for the economy and for society as a while.

## 4   Assessing the risk from natural hazards

The assessment of risk is a standard approach to the mitigation of severe natural hazards. It enables the community at risk to estimate the impact of that hazard long before it occurs and thus to consider what actions to take to handle that impact. These may include setting standards such that new developments can withstand the hazard, developing systems to reduce the impact of hazard on existing developments, preparation and validation of procedures to mitigate the impact when it arises. Such assessments are an important aspect of modern systems of governance; public authorities increasingly require risk assessment for wide range of developments, especially among public and private infrastructure. They will often require that new developments are designed to withstand local hazards up to some level with the 1 in 100-year risk typically being the lowest acceptable level. They will usually require higher standards for design of critical infrastructure, e.g. nuclear reactors, where the standard may be the 1 in 1000-year risk level. Thus the assessment of risk is central managing natural hazards. These assessments are critically underpinned by scientific knowledge of natural hazards and provide the means by which that knowledge can drive design standards and thus be used to improve the protection offered to humans and the infrastructures on which they depend.

How can we apply these ideas in the domain of space weather? – in particular how can we quantify the risks posed by severe space weather events? This is far from easy as we have limited information on such events. To demonstrate why this is a problem we first look at how risk is assessed in another natural hazard domain – namely flooding due to high flow rates on streams and rivers. A key feature of flooding (as with many other well-known natural hazards) is that is a local, rather than a regional or global, phenomenon. Intense rainfall is spatially structured with peak amounts varying markedly on spatial scales of just a few kilometres. Thus adjacent river systems can collect very different amounts of water and thus experience very different flow rates in the same storm event. Thus the flow rates on different rivers are statistically independent – a feature that has helped the hydrology community to develop robust statistical methods for assessing risks. When assessing the probability of high flow rates on any particular watercourse, it is possible to combine historical flow rate data from a number of similar watercourses and treat this as a single set of statistically independent data. The data from each watercourse must be normalized to the target stream. The combined dataset covers $N = \sum_s n_s$ years, where $n_s$ is the number of years of data collected on watercourse s and the sum is taken over all watercourses. It can then be analysed to determine the distribution of peak flow rate against return time and thus estimate what is the likely peak flow rate in any particular return time T. For that estimate to be considered reliable the dataset should cover at 5T years, i.e. N > 5T. Thus if N > 500 years, the dataset is considered adequate to assess the 1-in-100 year risk. This would not be feasible if the analysis were limited to the target stream, but is made feasible by the use of statistically independent data from multiple streams.

## 5  A statistical approach to space weather risks

What happens when we apply this approach to space weather data? We have a number of long-term data series that can be analysed to determine the distribution of high values against return time. We have done this analysis for two long-term data series: (a) the geoeffective electric field in the solar wind, which we can derive at hourly intervals for 44 years using the OMNI2 solar wind dataset, and (b) the geomagnetic index aa, which is available for 141 years.

The results of these analyses are shown in figures 2 and 3 respectively. In both cases we have assigned each valid data value to one of a logarithmically-spaced set of bins covering the full range of data values. The number of values assigned to each bin is then a measure of the occurrence frequency in the range of values covered by the bin. We show this occurrence frequency by plotting data points giving the number of values in each bin as a function of the lower boundary of the bin. We now look in detail at each figure.

Figure 2 shows the occurrence frequency for the geoeffective electric field in the solar wind, i.e. $Ey = - v \times Bz$, where v is the solar wind velocity and Bz is the GSM z component of the interplanetary magnetic field. The values of Ey were calculated using the OMNI2 interplanetary medium dataset from the US National Space Science Center. This gives hourly values of v and Bz measured near the Earth from November 1963 to the present. The data are necessarily taken from a variety of spacecraft and NSSDC has put considerable effort into cross-comparison and normalization of measurements from those different sources. It is the best available long-term dataset on the state of the solar wind. OMNI2 contains many data gaps due to a variety of problems including lack of downlink as well as instrument problems. These gaps can be different for plasma data (v) and for magnetic field data (Bz), so we check separately for the availability of valid values of v and Bz – and derive Ey only when both inputs are valid. We also assume v is radial from the Sun, so $Ey = -vBz$ and consider only the cases with $Bz < 0$, i.e. those that are significantly geoeffective. We have also applied a lower cut-off to the distribution corresponding to the Ey arising from $v = 300$ km s$^{-1}$ and $Bz = -0.1$ nT. The distribution below this cut-off is poorly sampled because the logarithmic bins become very small. These low values contribute nothing to our understanding of the occurrence of high values, so can be neglected. This gives a total of 120822 valid values of Ey over the period from 27 November 1963 to 31 December 2008. Thus we have valid data for 31% of the possible samples; this should be compared with a theoretical maximum of 50% (given the requirement that $Bz < 0$).

Figure 2 shows a surprisingly smooth distribution peaking around $Ey = 1$ mV m$^{-1}$, which corresponds to a potential of 100 kV when applied over the 100000 km scale size of the magnetosphere. This is consistent with the cross-cap potential often observed in the polar ionosphere in moderately disturbed conditions (e.g. through plasma flow measurements with SuperDARN) and which theory suggests is a separate measure of the solar wind potential applied to the magnetosphere. Above the peak, the distribution tends to a power-law as shown by the sloping line. This is a power law fit to the top 15 points in the distribution ($Ey > 2.4$ mV m$^{-1}$) and has a spectral index of -2.9. The two horizontal lines in Figure 2 indicate occurrence frequencies of 1 per 11 years (i.e. the average solar cycle duration) and of 1 per century. In principle we can use the intersection of the power law with these lines to estimate the maximum value of Ey expected per solar cycle and per century. However, you can immediately see that these would be poor estimates that cannot be relied on. The maximum value for a solar cycle is heavily determined by just two large data values (out of 120822) while the centennial maximum is an extrapolation well beyond all available data. These maxima are based on a mathematical fit to the data with little regard to physics, e.g. they provide no insight into physical factors that might limit the maximum value of Ey.

Figure 3 shows the occurrence frequency for the geomagnetic index aa. These were taken from the UK Solar System Data Centre and are available as a continuous series of three-hourly values from January 1868 to the present. It is the best available long-term dataset on the state of the occurrence of geomagnetic activity. This index is available in both final and provisional forms; we used the final form whenever available, and the provisional form when only that was available. This gives a total of 413528 valid values of aa over the period from 1 January 1868 to 11 July 2009. Thus we have valid data for 100% of the possible samples.

Figure 3 shows a noisy distribution at low values (aa < 15) but is smoother at higher values. The noisiness at low values is probably an artefact of the logarithmic bin sizes as discussed above – and, as in that discussion, has little relevance to our understanding of the occurrence frequency of high values. The overall distribution lacks a clear peak, is flat-topped at low values and shows a linear decrease at high values. Its form is suggestive of a Lorentzian distribution and the dashed line shows how such a distribution can be fitted to the data. This fit suggests that the high values follow a power law with spectral index -2.7. As in the previous figure the two horizontal lines indicate occurrence frequencies of 1 per 11 years and of 1 per century – and thus can provide an estimate of the maximum aa value expected on those timescale. And, as before, you can see that these maxima would be poor estimates that cannot be relied on.

The fundamental issue we face is that space weather is a global phenomenon, so measurements of relevant parameters on and around the Earth (such as Ey and aa, but also many others) are statistically correlated. Thus we cannot combine them to synthesise very long data series as is done in the assessment of other natural hazards. Thus the statistical approach is constrained by the limited duration of solar-terrestrial datasets. We really need 500 years of observations if we wish to assess the 1-in-100 years risk, but most of the STP datasets are much shorter than that. Among those datasets that are closely linked to geomagnetic activity we currently have 45 years of solar wind data, 78 years of ionospheric data and 140 to 160 years of geomagnetic data. These all fall far short of the required duration. The sunspot number data series is longer ranging from 180 years at daily resolution to 400 years for annual means. This is a useful indicator of the general level of solar activity but is not a measure of space weather events at the Earth.

## 6 Other approaches to space weather risks

### 6.1 Proxy data

One way to build longer time series datasets is to look for proxy data – something in the terrestrial environment whose attributes reflect the conditions prevailing when that thing was formed. One relevant example is the production of cosmogenic isotopes by the interaction of energetic particle radiation with the atmosphere. If those isotopes are trapped in an independently dateable deposit (e.g. annual ice layers in the permanent ice-sheets), they can be used to estimate the radiation fluence impinging on the Earth in the year of deposit. This technique has been used to estimate radiation fluences over the past several hundred years and continues to provide insights into the occurrence of major radiation events (Shea et al, 2006). Thus proxy data are becoming a powerful tool for assessing the historical occurrence of high fluence radiation events. However, we should note that they do not provide information on short-live very high flux events such as that on 23 February 1956, as discussed above.

*6.2 Space weather in other solar systems*

However, as yet, no proxy has been identified that can provide information on the historical levels of geomagnetic activity. Thus we have to find other methods for assessing the many aspects of space weather that arise from geomagnetic activity. One possibility here is to try to observe space weather in similar solar systems. If that were possible, we would have statistically independent measurements of space weather and could begin to follow the statistical approach used to study local natural hazards such as flooding. In essence we would have put space weather in a larger context where it becomes a local phenomenon. This makes the approach attractive but very challenging. The target solar systems must have a single star that is similar to the Sun, i.e. spectral type G, a significant convection zone in the outer part of the star in order to provide a dynamo for generation of magnetic fields, and a similar rotation rate to drive that dynamo. There are a number of such systems within 100 light-years of Earth and advances in astronomical instrumentation are opening up possibilities to observe space weather phenomena in those systems. For example, stellar flares have long been observed using space-based X-ray instruments. However, much astronomical observing has focused on more powerful objects and thus there is limited knowledge of flares on G-type stars. But there is little to stop the development of a suitable observing programme other than the need to obtain adequate telescope time.

Another space weather phenomenon that we might observe in other solar systems is the cyclotron maser emission from aurora on magnetised planets in those solar systems. This is the phenomenon known as auroral kilometric radiation at Earth and its intensity responds to the level of space activity in the magnetosphere. There is already considerable interest in observing cyclotron maser emission from exo-planets as it will provide a means to identify magnetised exo-planets and thus worlds that have some protection from cosmic radiation and from atmospheric erosion by stellar winds. Work in this area suggests that cyclotron maser emission from exo-planets will be observable by the next generation of radio telescopes such as LOFAR and SKA (Zarka, 2007). However, the frequency cut-off imposed by Earth's ionosphere will probably limit observations to larger-exo-planets with emissions in the MHz range. Observations of cyclotron maser emissions from Earth-like planets will probably require deployment of a sensitive low-frequency radio telescope in space or on the Moon (Lockwood, 2007).

Thus there is some potential to study space weather in other solar systems – and to use those observations to improve our statistical assessment of extreme space weather events. However, this is necessarily a long-term project as the techniques are in their infancy and, even when they are mature, it will take some years to build up an adequate dataset. This is a technique that will come to fruition in the middle of the $21^{st}$ century.

*6.3 Physics-based modelling*

The previous sections have shown that we have limited opportunities to assess the risks of extreme space weather via statistical analysis of observations. We therefore consider whether the modelling of space weather can provide an alternative means of assessing risk. In principle, we can use modelling to explore a range of space weather scenarios from quiet to moderate to extreme conditions and thereby accumulate information that would require centuries if we had to wait for nature to provide examples. However, this approach will work only if the modelling properly describes extreme conditions. This is essentially a requirement for the models to be physics-based and, in particular, to capture the physics at work in extreme events. We cannot use numerical models for this purpose as they are reliable only under average conditions (e.g. key inputs within a few standard deviations of the mean) and often generate bizarre results when presented with extreme conditions, far beyond those used to build the model. These results do not indicate a failure of the numerical model, but rather its inappropriate use.

Thus we seek to identify some of the key physics at work in extreme events and thereby identify areas that should be key targets for research on space weather models. In normal conditions, the magnetosphere mitigates the space weather effects induced by the solar wind and by cosmic radiation. These effects are reduced as they propagate into the closed magnetic field lines that envelope most of the Earth's surface, atmosphere and near-Earth space. Only the polar regions, where the magnetic field lines are usually open to the interplanetary medium, are exposed to the worst effects. During severe geomagnetic storms, there is a huge increase in the amount of open magnetic flux, so that the magnetic field lines at mid, or even low, latitudes become open to the solar wind. This is clearly demonstrated by the huge expansion of the auroral oval during severe storms. It is this expansion that generates many adverse space weather effects. The auroral electrojet will be moved to mid-latitudes greatly increasing the risk of geomagnetically induced currents in those regions. The presence of aurora at mid-latitudes will also have adverse effects on trans-ionospheric radio propagation, e.g. particle precipitation will alter the structure of the E region ionosphere, increased turbulence will generate strong scintillation effects. Auroral heating will change the density and composition of the mid-latitude thermosphere leading to changes to spacecraft drag and in the structure of the F region ionosphere.

An expansion of the region of open flux cannot be achieved just by increasing the speed and density of the solar wind, i.e. higher ram pressure. That pressure will just compress the magnetosphere; it will not increase the open flux. It will increase the level of activity in polar regions, but the magnetosphere will still reduce the impact of that activity as it propagates equatorward. Thus we need to consider the processes that control the open flux in the Earth's magnetosphere. Let us consider a simple model in which the amount of open flux is determined by the competition between (a) reconnection on the dayside magnetopause, which creates open flux, and (b) reconnection in the tail plasmasheet, which removes open flux. The rate of change of open flux is then given by the balance between these two reconnection processes and may be summarised by the following equation:

$$\frac{\partial \Phi_{open}}{\partial t} = \frac{\partial \Phi_{dayside}}{\partial t} - \frac{\partial \Phi_{tail}}{\partial t} \tag{1}$$

To open up the magnetosphere, as observed in severe geomagnetic storms, we need the first term on the right to be large and the second to be weak. The first term, the dayside reconnection rate, is equivalent to the potential applied to the magnetosphere by the convective electric field in the solar wind, i.e.

$$\frac{\partial \Phi_{dayside}}{\partial t} = -v_{SW} B_z L \qquad (2)$$

where $v_{sw}$ is the solar wind speed, Bz is the north-south component of the interplanetary magnetic field and L is the magnetospheric scale size over which the solar wind electric field is applied. The magnetospheric scale size is fixed (~100000 km), so to make this first term very large, we require a high solar wind speed $v_{sw}$ and a magnetic field component Bz that is strongly southward. Experience shows that the credible range of values includes $v_{sw}$ in excess of 2000 km s$^{-1}$ and Bz below -50 nT. This is sufficient to produce very high dayside reconnection with the solar wind applying a potential in excess of 10 million volts. If this were not balanced by tail reconnection, it would rapidly expand the open flux (as we shall discuss below) and produce the effects reported during severe geomagnetic storms. But for this mechanism to be effective, the tail reconnection rate must remain low for a sufficient time (perhaps five to ten minutes) to allow strong dayside reconnection to expand the region of open flux to mid-latitudes. Note that once the region of open flux has been expanded, it can remain large if dayside reconnection continues and balances any increases in tail reconnection rates as the storm proceeds. It will shrink only when dayside reconnection decreases to levels that allow tail reconnection to dominate.

Thus to understand how the region of open flux might be expanded, we must consider what process might act, during a severe storm, to slow tail reconnection without slowing dayside reconnection. There are several factors to consider:

- First, and most obvious, there is a natural time delay between the onset of dayside reconnection and the subsequent onset of tail reconnection. The solar wind electric field drives the transport of open flux to the tail. Thus tail reconnection will commence only after a time delay that can be no smaller than the flux transport time from the dayside. If the flux is transported at the solar wind speed, this time T is of order (Rd + Rt)/ V$_{sw}$, where Rd and Rt are the distances of the dayside and tail reconnection regions from Earth. If we take Rd = Rt = 10 Re, we find T is around 5 minutes for normal solar wind speeds of 400 km s$^{-1}$, decreasing to 1 minute for a high solar wind speed of 2000 km s$^{-1}$. This is too small to allow the required expansion of open flux.
- However, the transport of open flux may be slowed significantly below the solar wind speed. The solar wind electric field will generate Hall and Pedersen currents at the foot of open flux tubes and these currents will dissipate energy due to the finite conductivity in that region. This dissipation will slow flux transport and will be particularly marked during severe events. If we assume that the solar wind velocity maps down field lines to the ionosphere as the inverse square of the magnetic field strength, the example above ($v_{sw}$ > 2000 km s$^{-1}$ and Bz < -50 nT) would imply a plasma flow velocity > 60 km s$^{-1}$ in the ionosphere. This is an order of magnitude greater than the highest velocities usually seen in severe events. It is likely that ionospheric dissipation would significantly slow flux transport to the tail in severe events and thus increase the time delay before the onset of tail reconnection. A more detailed assessment of this process is beyond the aims of this paper, but it appears to be a mechanism that could operate in severe geomagnetic storms and one that could create the time delay needed to enable dayside reconnection to expand the region of open flux to mid-latitudes. Incidentally this dissipation will cause severe heating of the upper atmosphere and many space weather effects will follow from that.

- We should also consider if there are any mechanisms that could directly interfere with the tail reconnection process and reduce the effectiveness with which it destroys open flux. Efficient reconnection requires that plasma on newly reconnected flux tubes can freely flow away from the reconnection region. If not, the pressure from this outflow plasma will gradually slow and eventually halt the reconnection process. Thus we should consider if there are any processes that can choke the outflow from tail reconnection and that might come into prominence during severe storms. It is doubtful that there will be any changes in magnetic topology that could choke the outflow. But it is possible that increased mass loading of the tail plasma could have a similar effect. The onset of dayside reconnection will quickly be followed by strong geomagnetic activity in the ionospheric cusp region and this is likely to include strong outflows of ionospheric ions, in particular $O^+$ ions. The penetration of the solar wind electric field into the magnetosphere will quickly transport those ions to the tail and keep them from flowing back to the dayside reconnection region. Thus it is possible that the plasma in the tail reconnection region will quickly be loaded with extra mass. The increased mass density is likely to slow the reconnection rate since a given energy release by reconnection must then accelerate the outflow plasma to lower velocities.

Thus there are several possible mechanisms that by which tail reconnection might be slowed in severe space weather events. Further work is needed to quantify these mechanisms, but they support the idea that tail reconnection can be slowed sufficiently that dayside reconnection will dominate in early phase of a severe geomagnetic storm. If so, how long would it take to create the highly expanded region of open flux that appears to be a characteristic of severe geomagnetic storms? We can estimate this using a dipole model of the Earth's magnetic field. Figure 4 shows the amount of magnetic flux poleward of any latitude for a dipole of strength $7.64 \times 10^{22}$ Am$^{-2}$ (this is the dipole strength corresponding to the primary term ($g_1^0$) in Version 10 of the International Geomagnetic Reference Field, see http://www.ngdc.noaa.gov/IAGA/vmod/igrf.html). Figure 4 shows that, away from the polar region, the variation of magnetic flux with latitude is almost linear with $10^8$ webers of flux per degree of latitude. Thus, if a severe space weather event can generate a dayside reconnection voltage of 10 million volts as discussed above, we would expect the region of open flux to grow by 0.1° every second. Thus an expansion of the open flux boundary from 70° (its typical location at moderate activity) to 45° would take just over four minutes. This would expand the open flux to envelope all of Northern Europe and much of the North America and put a wide range of services at greatly increased risk from space weather. Further expansion would quickly (a few minutes) increase the risks in the many other populous regions closer to the equator.

We conclude that the rapid expansion of the open flux in Earth's magnetosphere is one of the key factors in determining the risks from severe space weather events, in particular for the many impacts that arise from severe geomagnetic activity. Thus to advance our understanding of these risk, we need to better understand the physical processes that will control the expansion of open flux in severe events. These include:
- Better understanding of the occurrence of the relevant conditions in the interplanetary medium, i.e. high solar wind speeds and strong southward magnetic fields.
- The extent to which dissipation in the polar ionosphere can slow the tailward transport of newly reconnected flux tubes, in particular the extent to which this might be enhanced during severe conditions.
- The extent to which ion outflows from Earth's upper atmosphere can modify tail dynamics, in particular slowing tail reconnection.

We need to understand the physics behind these processes and to incorporate that knowledge in space weather models so that they can be used to better simulate severe events and inform risk assessments.

## 7    Summary and conclusions

There is an urgent need to develop methods for quantifying the risks from severe space weather. Recent scientific work (Space Studies Board, 2008) suggests that the maximum credible risk from severe space weather would be very damaging for our modern technological civilisation. This risk has emerged in the past forty or fifty years as society has become increasingly dependent on services that exploit space-based infrastructure and long-distance transmission of electricity.

However, severe space weather falls in the category of low frequency high impact risks. These are rare events that only occur once in many decades, but with catastrophic consequences. Their low frequency presents a challenge to most systems of governance since these usually focus on immediate problems and pay attention to longer-term problems only when the consequences are made clear. Thus it is vital to raise awareness among policy-makers about the consequences of severe space weather events and of the likelihood of such events occurring in any particular time frame. The recent scientific advances in describing the consequences of a severe space weather event have raised public awareness about the problem, e.g. through articles in popular science magazines. However, much of that public awareness has focused on the specific risk at the next solar maximum – and lacks the scientific perspective that this is a risk that will continue forever but modulated by changes in solar activity.

It is therefore important to quantify the risks from severe space weather events and, in particular, to develop risk assessment methodologies that parallel those used to assess other natural hazards. This should produce quantitative results that will be meaningful to policy-makers. As noted by Fisher (2009) it is essential that space weather know-how is presented to policy-makers in ways that are credible, salient and legitimate. We argue here that a natural hazards approach to space weather risks (i.e. establishing robust estimates of occurrence frequency of high levels of space weather activity) is a critical step in that process.

This paper explores the statistical approach to natural hazards, using flood risk assessment as an exemplar of that approach. We show that it is mathematically possible to estimate the occurrence frequency of high levels of some key space weather parameters (specifically the convective electric field in the solar wind and the geomagnetic index aa). However, these estimates are not robust even at the level of 1-in-100 year risks because current STP datasets are far too short. Such estimates require time series measurements with an effective duration that is more than 500 years. For many other natural hazards, these time series can be synthesised by combining shorter time series from statistically independent instances of similar hazards. But that is not currently possible for space weather. We have substantial data series only for a single instance of space weather, namely solar activity on the earthward face of the Sun and its consequences on or near the Earth. These datasets are highly correlated, so we are limited to the duration for which relevant measurements are available, ranging from 45 years for conditions in the interplanetary medium to 140 to 170 years for geomagnetic measurements. If we confine our interest to such measurements we will have to wait several centuries to accumulate sufficient data.

We therefore examine alternative methods for making robust estimates of risk, assess their potential and suggest what future work is needed to deliver that potential. One alternative method is the use of proxy data. This is already an established approach for estimating historical radiation environment. It is already delivering important results, e.g. suggesting that the Carrington space weather event also included the outstanding solar radiation storm of the past several centuries. We can expect that work in this area will continue and is likely to provide further important results. However, there is as yet no proven method for using proxy data to quantify the strength of historical geomagnetic storms. It is possible that historical records may provide information on the latitudinal extent of major storms, but such records can be biased by a variety of factors including human error, weather and partial loss of records. An environmental proxy would enable systematic studies, but, as yet, no suitable feature has been identified.

A more speculative alternative method is to observe space weather phenomena in other solar systems similar to our own. This would give us access to statistically independent instances of space weather and thus open up the possibility to synthesise long time series by combining data from those instances. However, this is a challenging requirement at the present time. We can observe flares on other stars using modern X-ray telescopes but a systematic flare survey of Sun-like stars would require large amounts of time on those telescopes. Another technique that we must consider is to observe cyclotron master emissions from magnetised planets in similar solar systems. These are very strong radio emissions generated by the electrons that produce the aurora seen on magnetised planets so are well-correlated with the level of geomagnetic activity. Thus measurements of the strength of these emissions will provide a measure of the level of that activity – and one that can easily be compared with activity at the Earth. There is already considerable scientific interest in these emissions as they provide a way of detecting magnetised exo-planets. This has shown that the detection of these emissions should be within reach of radio telescopes within a decade or two – and also that it would be better to make such measurements in space so that we can observe the low frequency (<1 MHz) emissions from Earth-like exo-planets. We conclude that the observation of space weather phenomena in similar solar systems is an area that is ripe for development and has significant potential to help us assess the extremes of geomagnetic activity. It is an area that will grow in importance in the coming decades.

Another important alternative method is the use of physics-based modelling. If we could build space weather models that can reliably simulate extreme conditions, we would be able to use the simulations to assess the risks from those conditions. This approach offers the great advantage that we do not have to wait decades or even centuries to collect an adequate dataset. The challenge is to be confident that we have a reliable simulation of extreme conditions. Numerical models cannot be used as the applicability of any such model is inextricably tied to the average conditions within the dataset used to derive the model. For the reasons discussed throughout this paper, this cannot include an adequate sample of extreme conditions. It is therefore vital to use physics-based models and, furthermore, models that capture the key physics at work in extreme conditions. What is this key physics? We argue here that a major increase in open flux in Earth's magnetosphere is a key feature of severe space weather events at Earth. This expansion of open flux brings strong space weather effects to mid-latitudes and thus greatly increases the space weather risk in many heavily-populated regions. It causes bright aurora to be visible in these regions – a phenomenon that is well-attested in historical records. This expansion would require a short period of very strong dayside reconnection with little or no tail reconnection. In extreme solar wind conditions, this period could be quite short – just four minutes could be enough to expand the open-closed boundary from 70° to 45° latitude. Once expanded, all that is required to maintain the region of open flux is a balance between dayside and tail reconnection; this region will shrink only when dayside reconnection stops, thus allowing tail reconnection to remove open flux. Thus the key physics that we need to understand includes: (a) the occurrence of solar wind conditions that drive strong dayside reconnection (i.e. high solar wind speed and strongly southward IMF); and (b) the processes that can slow the onset of tail reconnection during the onset phase of a severe geomagnetic storm. This is an area where future research can give insights into the physics of extreme space weather. Key issues to consider include the effect of ionospheric dissipation on the tailward transport of newly opened magnetic flux and the modulation of tail reconnection rates due to mass-loading, e.g. $O^+$ outflows from the upper atmosphere.

In summary, this paper outlines several research approaches that can help us develop methods to quantify the risks from extreme space weather – in particular, to put those risks in a natural hazards framework, e.g. to estimate the occurrence rates for high levels of space weather activity. This quantitative information is a critical scientific input to the growing policy debate on the economic and societal impact of severe space weather. It will help policy-makers assess the priority that should be given to protection against severe space weather.

# 8 Acknowledgements

would like to thank the US National Research Council for supporting my participation in that workshop. I would also like to thank the US National Space Science Data Center for access to the OMNI solar wind dataset, the UK Solar System Data Centre for access to the aa and AE geomagnetic indices and the US National Geophysical Data Center for access to the aa*MAX dataset. I am also grateful for help from Ellen Clarke at British Geological Survey and Michel Menvielle at the International Service for Geomagnetic Indices, who confirmed the spurious aa values available for 10 May 2003 when this paper was prepared.

## Appendix A: aa*MAX

This index has been developed at the US National Geophysical Data Center as a way to assess the severity of geomagnetic storms. It is based on the well-known aa geomagnetic index, for which we have values back to 1868, but uses a running mean technique to provide a robust estimate of the strength and duration of each storm. The basic aa values are available at 3-hourly intervals. Thus individual data points are more indicative of the strength of individual sub-storms rather than the intensity of the overall storm. The aa index is also produced as daily values, i.e. the arithmetic mean of the eight 3-hourly values within a day as defined by Coordinated Universal Time and its predecessors. This provides a better estimate of the intensity of storm activity on a particular day, but is sensitive to the phasing of the storm. A storm that fills one UTC day will have a higher index than a similar storm that runs from mid-day to mid-day.

The aa*MAX index is intended to overcome this. It is based on an 8-point running mean (aa*) of the 3-hourly aa index, i.e. a 24 hour average that integrates over many sub-storms but is not constrained in phase. The duration of the storm is then taken as the period for which aa* is greater than 60, while the peak value of the running mean is aa*MAX, an estimate of overall strength of the storm.

Full details of aa*MAX, together with a complete set of values, are available on the NGDC website. At the time of writing, the set of values contains one wrongly identified storm, which has been ignored in this paper. This is an event listed as being on 10/11 May 2003 and ranked third highest in the values of aa*MAX. This event is spurious and arises from incorrect values of aa for 10 May 2003, which are now being corrected by the International Service for Geomagnetic Indices.

**Figure captions**

Figure 1. The variation of the four auroral electrojet indices before, during and after the severe geomagnetic storm on 13/14 March 1989.

Figure 2. Occurrence frequency for values of the geoeffective electric field in the solar wind using hourly data for 1963 to 2008. The frequency is shown as the count in each logarithmically space bin.

Figure 3. Occurrence frequency for values of the geomagnetic index aa using 3-hourly data from 1868 to 2009. The frequency is shown as the count in each logarithmically space bin.

Figure 4. The amount of magnetic flux poleward of any magnetic latitude for a magnetic dipole.of $7.64 \times 10^{22}$ Am$^{-2}$, i.e. similar to the Earth's dipole moment in 2005. The flux is shown in gigawebers.

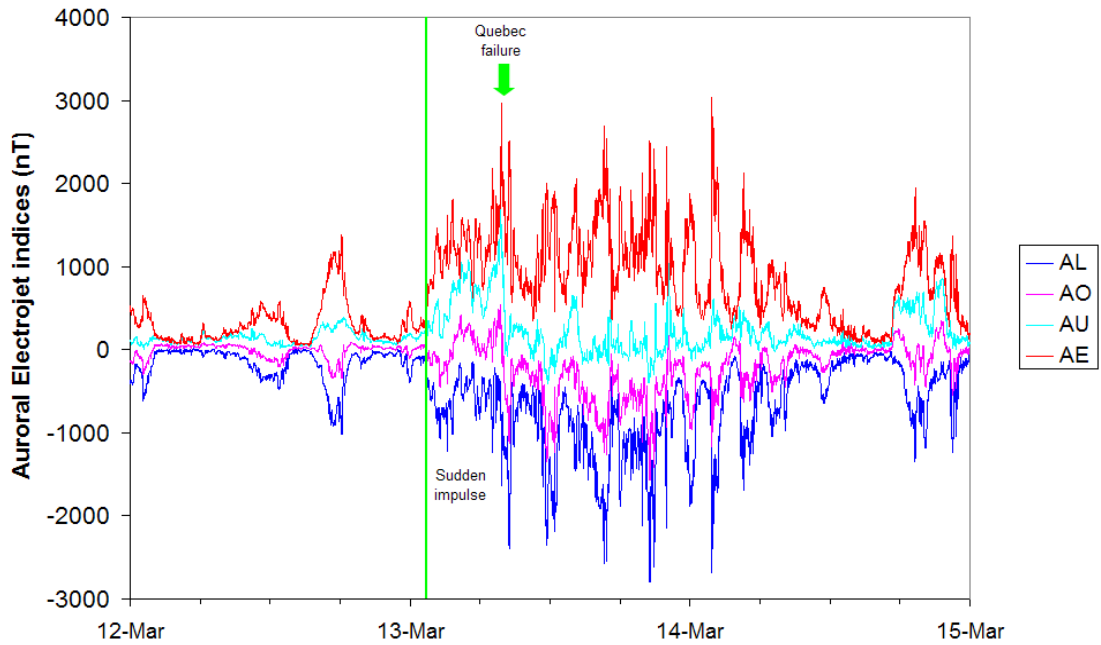

Figure 1. The variation of the four auroral electrojet indices before, during and after the severe geomagnetic storm on 13/14 March 1989.

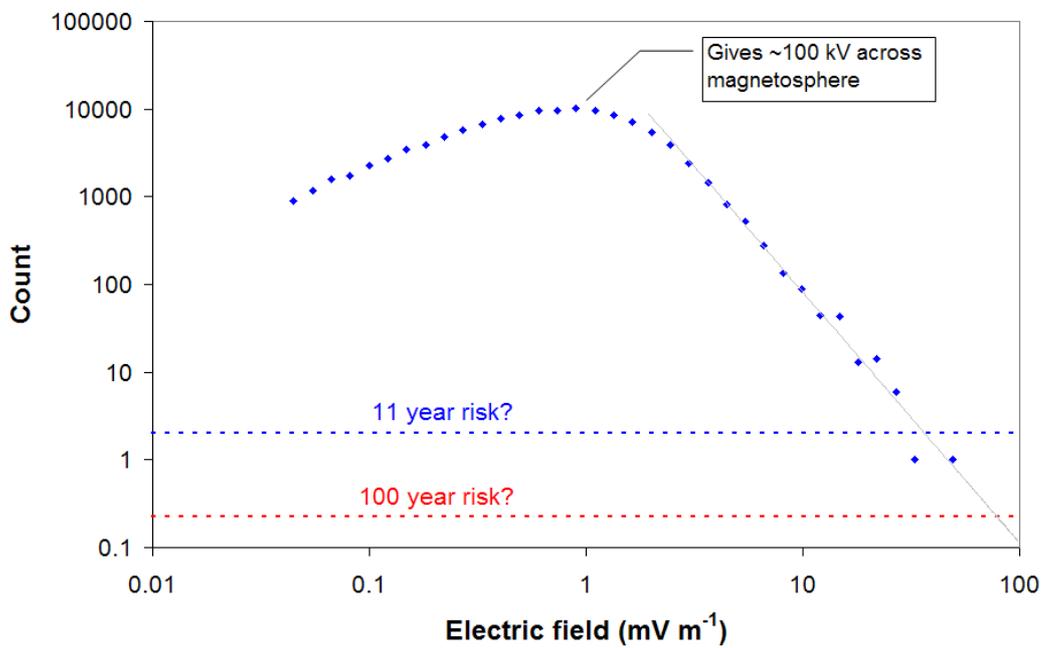

Figure 2. Occurrence frequency for values of the geoeffective electric field in the solar wind using hourly data for 1963 to 2008. The frequency is shown as the count in each logarithmically space bin.

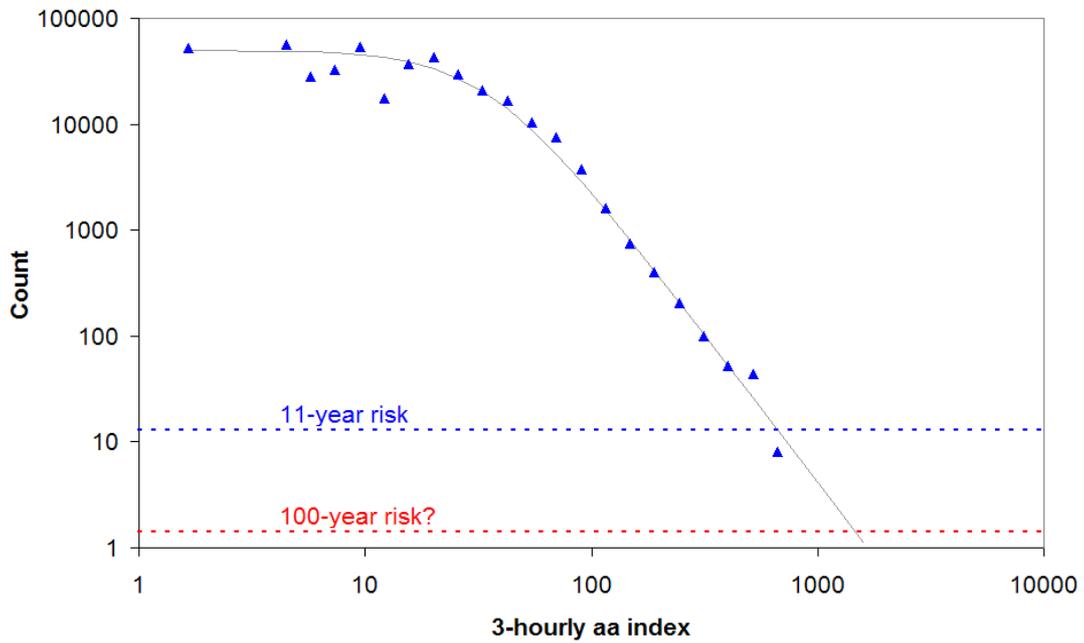

Figure 3. Occurrence frequency for values of the geomagnetic index aa using 3-hourly data from 1868 to 2009. The frequency is shown as the count in each logarithmically space bin.

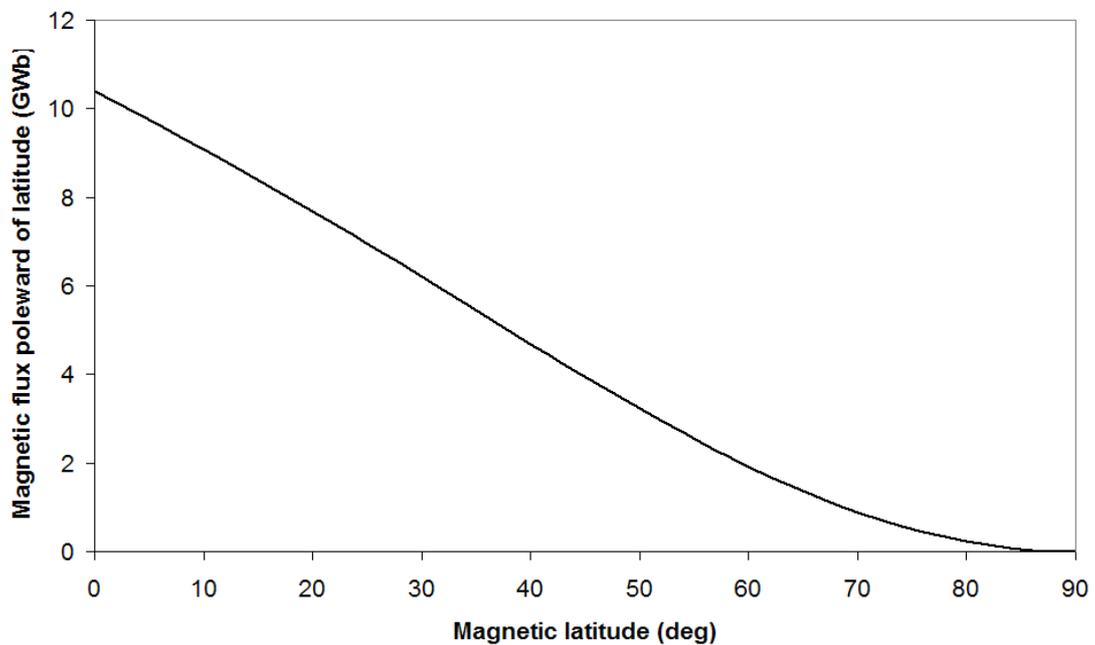

Figure 4. The amount of magnetic flux poleward of any magnetic latitude for a magnetic dipole.of $7.64 \times 10^{22}$ Am$^{-2}$, i.e. similar to the Earth's dipole moment in 2005. The flux is shown in gigawebers.